\documentclass[twocolumn,amsmath,amssymb,prl,10pt,nofootinbib,superscriptaddress]{revtex4}
\usepackage{ graphicx, float,amsmath, amsmath, amssymb}
\usepackage{xcolor} 
\usepackage[export]{adjustbox}
\usepackage[breaklinks, colorlinks, citecolor=blue]{hyperref}
\usepackage[labelsep=period]{caption}
\usepackage[normalem]{ulem}
\usepackage{mathtools}
\usepackage{siunitx}
\usepackage{soul}
\usepackage{physics}
\usepackage{minitoc}

\usepackage{bm}
\usepackage{xcolor}

\definecolor{alizarin}{rgb}{0.82, 0.1, 0.26}

\def\be{\begin{equation}}
\def\ee{\end{equation}}
\def\bea{\begin{eqnarray}}
\def\eea{\end{eqnarray}}
\def\bse{\begin{subequations}}
\def\ese{\end{subequations}}

\begin{document}
\setlength{\parindent}{0cm}

\title{Impact of generalized holonomy corrections on the cosmological primordial power spectra}

\author{Maxime De Sousa}%
\affiliation{%
Laboratoire de Physique Subatomique et de Cosmologie, Universit\'e Grenoble-Alpes, CNRS/IN2P3\\
53, avenue des Martyrs, 38026 Grenoble cedex, France
}

\author{Killian Martineau}%
\affiliation{%
Laboratoire de Physique Subatomique et de Cosmologie, Universit\'e Grenoble-Alpes, CNRS/IN2P3\\
53, avenue des Martyrs, 38026 Grenoble cedex, France
}

\author{Cyril Renevey}%
\affiliation{%
Laboratoire de Physique Subatomique et de Cosmologie, Universit\'e Grenoble-Alpes, CNRS/IN2P3\\
53, avenue des Martyrs, 38026 Grenoble cedex, France
}

\author{Aur\'{e}lien Barrau}%
\affiliation{%
Laboratoire de Physique Subatomique et de Cosmologie, Universit\'e Grenoble-Alpes, CNRS/IN2P3\\
53, avenue des Martyrs, 38026 Grenoble cedex, France
}



\date{\today}
\begin{abstract} 
The propagation of perturbations is studied with generalized holonomy corrections in a fully consistent way, ensuring that the deformed algebra of constraints remains closed. The primordial cosmological power spectra are calculated. It is shown that, although the detailed form of the correction does unavoidably impact the observables, the main known results of loop quantum cosmology are robust in this respect.
\end{abstract}
\maketitle

\section{Introduction}

Loop quantum gravity (LQG) is a background-independent quantization of general relativity (GR) \cite{Ashtekar:2021kfp}. It can be expressed in the canonical form \cite{Rovelli:2004tv} or in the covariant way \cite{Rovelli:2014ssa}. The theory has been successfully applied to both black holes and the early universe. Many consistency checks -- mostly encouraging -- have been carried out, although some important questions remain open \cite{Kiefer:2007gns}. An excellent philosophical introduction is given in \cite{Rovelli:2022xsj}.\\

The cosmological sector of the theory has received a particular attention and numerous complementary aspects were investigated in details (see, {\it e.g.} \cite{Ashtekar:2007em,lqc9,Agullo2,Barrau:2013ula,Diener:2014mia,Ashtekar:2015dja,Bolliet:2015bka,Alesci:2015nja,Martineau:2017sti,Gielen:2017eco}, and references therein). The main conclusions are the following. The existence of a bounce replacing the usual big bang is a robust result. It has been shown analytically in simplified situations and proven to survive when a cosmological constant is added, when spatial curvature is taken into account, and when quite general potentials for the inflaton field are considered. In addition, the semiclassical states were demonstrated to remain sharply peaked, allowing the safe use of effective equations. Importantly, the duration of inflation is statistically predictable in this framework. Generic features for the primordial power spectra were also derived.\\

This work deals with a specific -- and somehow underestimated -- point:
the consequences of a generalized holonomy correction. The outstanding issue of quantization ambiguities in LQG was mentioned in \cite{Perez:2005fn}.  New arguments were recently given in \cite{Amadei:2022zwp}. In particular, the question was addressed from the interesting point of view of renormalization. The quantization ambiguity of the connection-based holonomy variable might influence the associated cosmological  predictions. This  has been studied in \cite{Renevey:2021tmh}. The main effects are quite weak on the background dynamics and do not change substantially the usual conclusions of loop quantum cosmology (LQC).
Interestingly, most new effects tend to decrease the number of e-folds. This makes the situation more phenomenologically promising. Perturbations were also considered in this work. However, the usual Mukhanov-Sasaki equations for gauge-invariant perturbations were used, which are not fully consistent with the underlying deformed algebra. The effects of the holonomy modifications were accounted for at the level of the background and at the level of the potential, but not in the core of the propagation equation. This article fills this gap and shows the calculation of fully reliable primordial spectra (in the deformed algebra approach). The main conclusion is that the known results of LQC are robust.\\

In the first section, we review the basics of LQC so that this article is self-contained for nonspecialists. Then, the deformed algebra and the propagation equations for perturbations are defined. Finally, the results are shown for different  parametrizations of the holonomy correction.

Throughout all the article, we use Planck units.

\section{Basics of Loop Quantum Cosmology}

Loop quantum cosmology is an attempt to perform a symmetry reduction of LQG, mimicking the quantization used in the full theory \citep{Ashtekar:2011ni,Bojowald:2005epg}. This section explains the basic ideas for the unfamiliar reader.\\

The canonical formulation of LQG is based on the Ashtekar connection,

\begin{equation}
    A^i_a := \Gamma^i_a + \gamma K^i_a,
\end{equation}

where $\gamma$ is the Barbero-Immirzi  parameter and the extrinsic curvature coefficients are given by $K^i_a=K_{ab} e^b_j \eta^{ij}$ for  triads defined such that $q^{ab}=e^a_i e^b_j \delta^{ i j}$ at each $x\in\Sigma_t$. The spin connection $\Gamma^i_a$ reads

\begin{equation}
    \Gamma^i_a = -\frac12 \epsilon^{ij}_{\ \ k} e^b_j \biggl( \partial_{[a}e^k_{b]} + \delta^{kl}\delta_{ms} e^c_l e^m_a \partial_b e^s_c \biggr),
\end{equation}

$e^a_i$ being the inverse triads such that $e^a_i e_b^j = \delta_i^j$. In order to complete the set of canonical variables, one defines the densitized triads $E^a_i:=|\textrm{det}\ {e}|^{-1} e^a_i$ that are conjugate to the Ashtekar connection,

\begin{equation}
    \left\{ A^i_a(x), E^b_j(y)\right\}=\kappa \gamma \delta_a^b \delta^i_j \delta^3(x-y),
    \label{eq:commu}
\end{equation}

with, $\kappa=8 \pi G$. \\

The dynamical equations appear as constraints. Namely, the Gauss constraint,

\begin{equation}
    G[\Lambda] = \left( \kappa \gamma \right)^{-1} \! \! \int_{\Sigma_t} \! \! \! d^3 x \Lambda^i \left( \partial_a E^a_i + \epsilon^\ell_{ik} A^k_a E^a_\ell \right),
\end{equation}

the diffeomorphism constraint, 

\begin{equation}
    D \left[N^a\right] = \left( \kappa \gamma \right)^{-1} \! \! \int_{\Sigma_t} \! \! \! d^3 x N^a F^i_{ab} E^b_i,
\end{equation}

where $F^i_{ab} = 2 \partial_{[a}A^i_{b]} + \epsilon^{i}_{jk} A_a^j A_b^k$ is defined as the curvature of the Ashtekar connection, and the scalar constraint,

\begin{gather}
    C\left[ N \right] = \left( 2\kappa \gamma \right)^{-1} \! \! \int_{\Sigma_t} \! \! \! d^3 x N|\textrm{det} \ E|^{-1/2} \Bigl( \epsilon_{ijk} F^i_{ab} E^a_j E^b_k \notag \\ \qquad \qquad\qquad\qquad - 2 \left(1+\gamma^2\right)K^i_{[a} K^j_{b]}E^a_i E^b_j \Bigr).
\end{gather}

For an isotropic, homogeneous and flat universe, the Friedmann-Lemaître-Robertson-Walker (FLRW) metric can be written as

\begin{equation}
    ds^2 = -N^2 \left( dx^0 \right)^2 + a^2(t)dx^a dx^b \delta_{ab}.
\end{equation}

The Ashtekar variables are rewritten as:

\begin{equation}
    A^i_a(x) = \gamma \dot{a}(t) \delta^i_a \equiv c(t) \delta^i_a, \quad 
    E^a_i(x) = a^2(t) \delta^a_i \equiv p(t) \delta^a_i,
\end{equation}

where a dot denotes a derivative with respect to the cosmic time $dt = N dx^0$. 
Only the scalar constraint contributes to the dynamics of this system. Taking into account the symmetries, it can be written as,

\begin{equation}
    C[N] = -\frac{3 N V_0}{\kappa \gamma^2} p^{1/2} c^2,
\end{equation}
where $V_0$ is a fiducial volume element.

The matter sector is assumed to be a scalar field $\phi$ with an arbitrary potential $V(\phi)$. The full Hamiltonian is

\begin{equation}
    H_t[N] = N V_0 \left( -\frac{3}{\kappa \gamma^2} p^{1/2} c^2 + p^{3/2} \rho \right),
\end{equation}

which, after setting $H_t[N] = 0$, leads to the usual Friedmann equation. 

The holonomy around the closed fiducial square $\Box_{ij}$ can be written as

\begin{equation}
    h_{\Box_{ij}}=h_{l_i}h_{l_j}h^{-1}_{l_i}h^{-1}_{l_j},
\end{equation}

with

\begin{equation}
    h_{l_i}=\exp \bigl\{|l| k \tau^i \bigr\},
\end{equation}

where $\tau^i$ are base matrices of the fundamental $SU(2)$ representation, which is arbitrary at this point. Hence, the holonomy-corrected curvature is 

\begin{equation}
    F^k_{ab}= -2 \lim_{l \rightarrow 0} \text{tr}\left[ \frac{h_{\Box_{ij}} - 1}{l^2} \tau^k \frac{e^i_a e^j_b}{\gamma^2} \right],
\end{equation}

which is equivalent to

\begin{equation}
     F^k_{ab}= \lim_{l \rightarrow 0} \frac{\sin^2\left( |l| k \right)}{|l|^2} \epsilon^k_{ij} \frac{e^i_a e^j_b}{\gamma^2}.
\end{equation}

The presence of a minimal area in the theory, given by the smallest nonzero eigenvalue of the area operator in LQG, leads to the introduction of the $\bar{\mu}=l_{Pl} \left( 4 \sqrt{3} \pi \gamma \right)^{1/2} p^{-1/2}$ parameter which allows to introduces the so-called holonomy correction substitution,

\begin{equation}
    c^2 \longrightarrow \bar{\mu}^{-2} \sin^2 \left( \bar{\mu} c \right).
    \label{Eq. mubar introduction}
\end{equation}

Gathering everything, the holonomy-corrected Hamiltonian constraint becomes

\begin{equation}
     H_t[N] = N V_0 \left( -\frac{3}{\kappa \gamma^2\bar{\mu}^{2}} p^{1/2} \sin^2 \left( \bar{\mu} c \right) + p^{3/2} \rho \right),
\end{equation}

which leads to the LQC-modified Friedmann equation,

\begin{equation}
    H^2 = \frac{\kappa}{3} \rho \left( 1- \frac{\rho}{\rho_c} \right),
\end{equation}
where $\rho_c = \sqrt{3} / (4 \pi \kappa \gamma^3)$  is usually assumed to be close to the Planck density. This is the usual bounce solution.\\

Many arguments (see, {\it e.g.} \citep{BenAchour:2016ajk,Yang:2009nj}) were given for considering seriously LQC with arbitrary spin representations or higher-order terms. In this work, we will remain as general as possible. To this aim, we will focus on a so-called \textit{polymerization} defined by the substitution,
 
 \begin{equation}
     c^2 \longrightarrow g^2(c, p),
 \end{equation}

 Details about the construction of suitable semiclassical states and associated Dirac observables are given in \cite{Ashtekar:2007em}.

where the only restriction on the (periodic) $g(c,p)$ function is the low-curvature limit, in which GR should be recovered, {\it i.e.} $g(c,p)\longrightarrow c$.\\

In order to set notations let us recall some known results for the polymerized background dynamics in LQC. As shown in \cite{Han:2017wmt}, the background dynamics is described by the set of equations

\begin{equation}
\begin{aligned}
\dot{c} &= -\frac{3N}{2\sqrt{p}} g^2(c,p) + \frac{Nk}{\sqrt{p}} \mathcal{G}^{(1)}(c,p) - N \frac{\kappa}{2} \sqrt{p} P, \\
   \dot{p} &= 2N \sqrt{p} \mathcal{G}^{(1)}(c,p), \\
   \dot{\phi} &= N \pi p^{-3/2}, \\
   \dot{\pi} &= - N p^{3/2} \partial_\phi V(\phi),
\end{aligned}
\label{eq:seteqdyn}
\end{equation}

where, as defined above, dots correspond to derivatives with respect to the cosmic time $t$, $\left\lbrace \phi, \pi \right \rbrace$ are the canonical variables for a given minimally coupled scalar field with potential $V(\phi)$ and pressure $P$. We also used the notation $\mathcal{G}^{(1)}(c,p):=\partial_c g^2(c,p)/2$.\\

The background Hamiltonian constraint can be rewritten as

\begin{equation}
    3 \sqrt{p} g^2(c,p) = \kappa \rho,
\end{equation}

where $\rho = \frac{\pi^2}{2p^3} + V(\phi) $. We make the usual gauge choice $N=1$, which allows us to rewrite the above set of equations as a generalized Friedmann equation, together with the usual Klein-Gordon equation for the inflation field:

\begin{equation}
\begin{aligned}
&H^2 = \frac{\kappa}{3} \rho \left( \partial_c g(c,p) \right)^2,\\
    &\ddot{\phi} + 3 H \dot{\phi} + \partial_\phi V(\phi) = 0,
\label{eq:dyn}
\end{aligned}
\end{equation}

where $H:=\frac12\dot{p}p^{-1}$ is the Hubble parameter.\\

For the background dynamics given above, four initial conditions are needed: the scale factor $a$, the Hubble parameter $H$, the scalar field $\phi$, and its time derivative $\partial_t \phi$, have to be determined at some specific time. The Ashtekar school has usually advocated the (reasonable) idea that the bounce time should be chosen whereas the Grenoble school prefers the pre-bounce classical universe. The dynamics at the bounce being dominated by quantum effects, we adopt this second choice (which is anyway meaningful if the bounce state is to be understood as the result of causal evolution from the contracting branch). In addition, in the prebounce phase, one can define a clear measure for probabilities \cite{bl,Linsefors:2014tna,Martineau:2017sti} relying on `safe" equations.\\

We impose $a(t_i)=1$. Far in the prebounce phase, the universe is mostly classical and the Hubble parameter is then given by the usual Friedmann equation, {\it i.e.} $H(t_i)=-\sqrt{\kappa \rho (t_i)/3}$. To discuss initial conditions for the matter sector, we introduce two parameters $x$ and $y$ defined by

\begin{equation}
\begin{aligned}
    &x=\bigg( V(\phi) / \rho_c \bigg)^{1/2},\\
    &y=\bigg( \dot{\phi}^2 / 2 \rho_c \bigg)^{1/2},
\end{aligned}
\end{equation}

satisfying the relation

\begin{equation}
    x^2 + y^2 = \frac{\rho}{\rho_c}.
\end{equation}

 In this study, we assume a quadratic potential (that is a simple mass term) for the field. Even though this potential is not favored by observational data \cite{Planck:2018jri}, it allows easy comparisons with other studies. Our results do not, in any case, significantly depend on the shape of the potential. We have explicitly checked this with the Starobinsky potential. In the remote contracting universe, the dynamics of $x$ and $y$ is described by a harmonic oscillator,

\begin{equation}
\begin{aligned}
    &x(t)\approx\left( \frac{\rho(t_i)}{\rho_c} \right)^{1/2} \sin \left( mt + \delta \right),\\
    &y(t)\approx\left( \frac{\rho(t_i)}{\rho_c} \right)^{1/2} \cos \left( mt + \delta \right),
\end{aligned}
\end{equation}

where the phase parameter $\delta$ has been studied in \cite{Linsefors:2013cd} and is not of particular importance in this work. The initial density is

\begin{equation}
    \rho(t_i=0)\approx\rho_c \left( \frac{\Gamma}{\alpha} \right)^2 \left[ 1- \left( 4 \alpha \right)^{-1} \sin \left( 2 \delta \right) \right],
\end{equation}

with $\Gamma$ the ratio of the classical timescale over the quantum one and $\alpha$ is a free parameter (set, as usually, to $\alpha=17 \pi + 1$ to ensure that the scalar field oscillates enough during the contracting phase for our approximations to be valid).

\section{Deformed algebra and perturbation equations} 

\subsection{The deformed algebra approach}

The treatment of perturbations is less consensual than the one of the background. On the one hand, the so-called \textit{dressed metric} approach (see \cite{Agullo:2012fc,Agullo:2013ai} for an introduction) was developed to account for quantum effects as deeply as possible. It is basically equivalent to the hybrid quantization one \cite{Li:2022evi} and the propagation equation is the usual one. On the other hand, the \textit{deformed algebra} framework (see \cite{Barrau:2014maa} for an introduction) was suggested to put emphasis on covariance. It is the main focus of this study as it constitutes the natural path to investigate the specific effects of the generalized holonomy correction in a self-contained way.\\

Basically, the deformed algebra approach is a conservative one which relies on consistency. In the canonical formulation of GR, the smeared constraints form a first-class algebra.
This closure property -- that is, the fact that each Poisson bracket between constraints is proportional to another constraint --  ensures that the evolution vectors always remain tangent to the submanifold of constraints. In other words, this makes the constraints compatible with themselves. When holonomy corrections are implemented, the resulting quantum gravity effective constraints do, however, not close anymore for perturbations (the closure is automatically ensured for the background). In the seminal work \cite{Bojowald:2008gz}, an elegant and consistent way out was found.  The interested reader can find details, {\it e.g.}, in \cite{tom1,tom2}. Important consequences were derived on the allowed shapes of the correction in \cite{eucl2}. To cancel the so-called anomalies, that is $\mathcal{A}_{IJ}$ terms appearing in Poisson brackets between (smeared) corrected constraints ${C}^Q_I$, 
\begin{equation}
\{ \mathcal{C}^Q_I, \mathcal{C}^Q_J \} = {f^K}_{IJ}(A^j_b,E^a_i) \mathcal{C}^Q_K+
\mathcal{A}_{IJ},
\end{equation}
one adds counterterms physically encoding the deformation of the algebra. A pictorial representation is given in \cite{tom1}. Those terms are required to vanish in the classical limit and are uniquely determined by the full system of equations (including matter). Quite amazingly, similar conclusions were reached in \citep{Han:2017wmt}, with a more general holonomy substitution. The fact that an anomaly-free algebra can still be constructed, always requiring the $\bar{\mu}$-scheme, is a strong hint in favor of the consistency of this path. \\

The idea, when considering linear perturbations, is to perturb constraints (and so the Hamiltonian) up to the quadratic order and to add counterterms (vanishing in the classical limit) to prevent anomalies. The Poisson brackets between all constraints are explicitly calculated. The calculations are quite involved but the final result is surprisingly elegant and simple:

\begin{eqnarray}
    \left\{G[\Lambda],G[\Lambda']\right\} = 0 ,\\ \left\{D_{tot}[N^a],G[\Lambda]\right\} = 0 ,\\
    \left\{H_{tot}[N],G[\Lambda]\right\} = 0 ,\\
\left\{ D_{tot}[N^a_1],D_{tot} [N^a_2] \right\} =0,
\end{eqnarray}

and 

\begin{equation}
\left\{ H_{tot}[N],D_{tot}[N^a] \right\} = - H_{tot}[\delta N^a \partial_a \delta N],
\end{equation}
together with
\begin{eqnarray}
\left\{ H_{tot}[N_1],H_{tot}[N_2] \right\} = 
\left(\frac{1}{2}\frac{\partial^2 g^2(c,p)}{\partial c^2}\right)\\
\times D_{tot} \left[  \frac{N}{p} \partial^a(\delta N_2 - \delta N_1)\right].
\label{HHbrac}
\end{eqnarray}

The factor $\frac{1}{2}\frac{\partial^2 g^2(c,p)}{\partial c^2}$ tends to 1 in the classical limit. When this factor becomes negative, the signature of spacetime changes to Euclidean, in agreement with what happens in the $\bar{\mu}$-scheme near the bounce. This has far-reaching consequences, from a specific phenomenology \cite{Mielczarek:2012pf,Mielczarek:2012tn,Mielczarek:2013rva,Linsefors:2012et,Barrau:2013ula,Barrau:2014wta,Mielczarek:2014kea,Ashtekar:2015dja,Schander:2015eja,Bolliet:2015bka,Barrau:2016sqp,Martineau:2017tdx,Barrau:2018gyz,Wilson-Ewing:2016yan} to unforeseen links with the Hartle-Hawking proposal \cite{Bojowald:2018gdt,Bojowald:2020kob}. 

A contradiction with data was noticed in \cite{Bolliet:2015raa} due to the power increase in the UV part of the spectrum, associated with the Euclidean phase. It is very important to underline -- as this point is often misunderstood -- that this result does {\it not} mean, in any way, that the deformed algebra approach to LQC is discarded. Just the other way around, it shows that this framework is   suited at making potentially testable predictions. It could very well be that the deformed algebra captures the main features of loop gravity and that LQC in itself is falsified. It could also be that the observational window does not fall in the altered part of the comoving spectrum (if inflation is brief). It could finally be that the way perturbations are propagated in the ``timeless phase" is incorrect, which has nothing to do with the deformed algebra framework itself \cite{Bojowald:2015gra}. It might even be that initial conditions are not properly set \cite{Mielczarek:2014kea}. \\

The main point that has to be underlined at this stage is that as long as the function $g^2(c,p)$ changes concavity, a signature change in unavoidable in this (conservative) approach, as mentioned in \citep{Han:2017wmt}. As $g^2(c,p)\sim c^2$ near the origin (to recover GR) and as the function is periodic, the change of concavity automatically happens. This is a strong conclusion. However, contrarily to what is written in \cite{Han:2017wmt}, this does not necessarily happen near the maximum of the function. Otherwise stated, for generalized holonomy corrections, the signature change is unavoidable but the energy density at which it takes place does not need to be close to the one of the bounce.\\

\subsection{Perturbation equations}
 
Quite a few results were derived both for the background and the perturbations in \cite{Renevey:2021tmh}. However, the equation for perturbations was not fully consistent. This is what we correct here.\\

The perturbed Einstein equations for a flat Universe filled with a scalar field in the polymerization framework has already been derived in the deformed algebra approach \cite{Han:2017wmt}. The Hamilton equation of motion for background variables is written thanks to the elementary Poisson brackets, as previously explained. Following the standard procedure, the equations of motions for the perturbed variables are decomposed in scalar, vector, and tensor modes. The physical part is then extracted by considering terms invariant under both Gauss and diffeomorphism transformations. This results in:
 
 \begin{equation}
     v''_{S/T} - \mathcal{G}^{(2)} (c,p)\nabla^2 v_{S/T}- \frac{z''_{S/T}}{z_{S/T}}v_{S/T} = 0,
     \label{eq:mukha}
 \end{equation}

where the prime  denotes a derivation with respect to the conformal time $d\eta=p^{-1/2}dt$ and $v_{S/T}$ is the Mukhanov variable for, respectively, scalar and tensor modes. We have also  defined $\mathcal{G}^{(2)}(c,p):=\partial_c \mathcal{G}^{(1)}(c,p)$. This in agreement with what was previously found for the usual correction \cite{eucl2}. \\

By performing a Fourier decomposition on the $k$ modes and introducing the variables $h_{k}=v_k/z$ and $\tilde{g}_k = \sqrt{p} \dot{h}_k / \mathcal{G}^{(2)}(c,p)$, it is possible to rewrite Eq. (\ref{eq:mukha}) as a set of first-order coupled differential equations. For tensor modes, one gets

\begin{equation}
\begin{aligned}
&\dot{h}_k = \frac{\mathcal{G}^{(2)}(c,p)}{\sqrt{p}} \tilde{g}_k, \\
&\dot{\tilde{g}}_k = -2H\tilde{g}_k - a k^2 h_k,
\end{aligned}
\end{equation}

whereas, for scalar modes, the equations are

\begin{equation}
    \begin{aligned}
        &\dot{h}_k = \frac{1}{\sqrt{p}} \tilde{g}_k, \\ 
        &\dot{\tilde{g}}_k = -2H \tilde{g}_k - a \mathcal{K}(k, t, c, p) h_k,
    \end{aligned}
\end{equation}

with $\mathcal{K}(k, t, c, p) =\mathcal{G}^{(2)}(c,p) k^2 a^{-2} - H \dot{z}_S z_S^{-1} - \ddot{z}_S z_S^{-1}$. \\

\subsection{Initial conditions for perturbations} 

Following the logic of causality (and remaining consistent with the background evolution), the initial conditions for perturbations are set in the prebounce contracting branch. The perturbations are thereafter propagated through the bounce and the Euclidean phase until they exit the horizon during the inflationary stage. This approach is different from the one depicted in \citep{Han:2017wmt} in which the authors set initial conditions for the perturbations at the onset of inflation. In this latter case, by construction, the perturbations never feel the high energy quantum regime and the Euclidean phase. This is why our results are deeply different.\\

The usual canonical quantization procedure is applied for each mode $v_k$. In the Heisenberg picture, 

\begin{equation}
    \hat{v}_k(\eta) = v_k(\eta) \hat{a}_k + v_k^*(\eta) \hat{a}^\dagger_{-k},
\end{equation}

where $\hat{a}_k, \hat{a}^\dagger_k$ are, respectively, the annihilation and creation operators, satisfying the usual commutation relation. This leads to the so-called Wronskian condition:

\begin{equation}
    v_k \frac{d v_k^*}{d\eta} - v_k^* \frac{d v_k}{d \eta} = i,
\end{equation}

implying restrictions for the mode coefficients. In particular, the Minkowski vacuum can be rewritten (in the case of tensors modes) as:

\begin{equation}
\begin{aligned}
&h_k(t_i) = \left(2k\right)^{-1/2}a^{-1}(t_i),\\
    &\tilde{g}_k(t_i) = -i\left( k/2 \right)^{-1/2}a^{-1}(t_i)  - \left(2k\right)^{-1/2} H(t_i).
\end{aligned}
\end{equation}

Initial conditions for scalar modes are way harder to derive, in particular due to the shape of the potential in the contracting branch of (all) bouncing models \cite{Barrau:2018gyz}. One can however rely on an appropriate WKB approximation \cite{Schander:2015eja}. In this approach, we constrain the mode coefficients from the Wronskian equation and choose the coefficients to describe a wave propagation in the positive time direction. One is then able to derive the initial conditions of $h_k$ and $g_k$ for scalar modes. 

\section{Numerical results and discussions}

For Gaussian perturbations,
the full statistical information is given by the 2-point correlation function. In a very standard manner, the scalar and tensor primordial power spectra are expressed as functions of the Mukhanov variable and the associated potentials evaluated at the horizon crossing:



\begin{equation}
	P_S(k)= \left. \frac{k^3}{2\pi^2} \left| \frac{v_k}{z_S} \right|^2 \right\vert_{k=aH} 
\end{equation}

and

\begin{equation}
   P_T(k) = \frac{4k^3}{\pi^2} \left. \left| \frac{v_k}{z_T} \right|^2 \right\vert_{k=aH}.
\end{equation}





In order to study the polymerization effects on those primordial power spectra, one needs to choose explicit expressions for the $g(k, p)$ function. There are not many constraints on the shape of $g$: mainly the low-energy limit and periodicity. Still, following \cite{Han:2017wmt}, solving the anomalies in the algebra of constraints imposes $g$ to be of the following form:

\begin{equation}
    g(c, p) = p^{1/2} \varphi (cp^{-1/2}),
\end{equation}

$\varphi (cp^{-1/2})$ being an arbitrary function of $(cp^{-1/2})$. For the rest of this work we rewrite

\begin{equation}
  g(c, p)  := \Bar{\mu}^{-1} f(x)
\end{equation}

where $\Bar{\mu}$ is the parameter already introduced Eq. (\ref{Eq. mubar introduction}). With this notation, one can easily retrieve the usual LQC prescription with $f(x)=\sin(x)$. 

Several functions have been considered in \citep{Renevey:2021tmh}, for example

\begin{equation}
    f_{sqr}(x)=\sin(x) \sqrt{1+A_1 x^{n_1} (x-\pi)^{n_2}}, 
\end{equation}

with $n_i \geq 1$, $i\in \left\{ 1,2 \right\}$ and $A_1\geq 0$, together with

\begin{equation}
f_{cos}(x) = \sin(x) \sqrt{\left( 1+C_1 \right)^{-1} \sum^{C_1}_{n=0} \cos^{2n}(x)},
\end{equation}

with $C_1 \geq 1$. One can easily show that such parametrizations have the correct behavior in the low-energy limit.

\begin{figure}[H] 
\includegraphics[width=\linewidth]{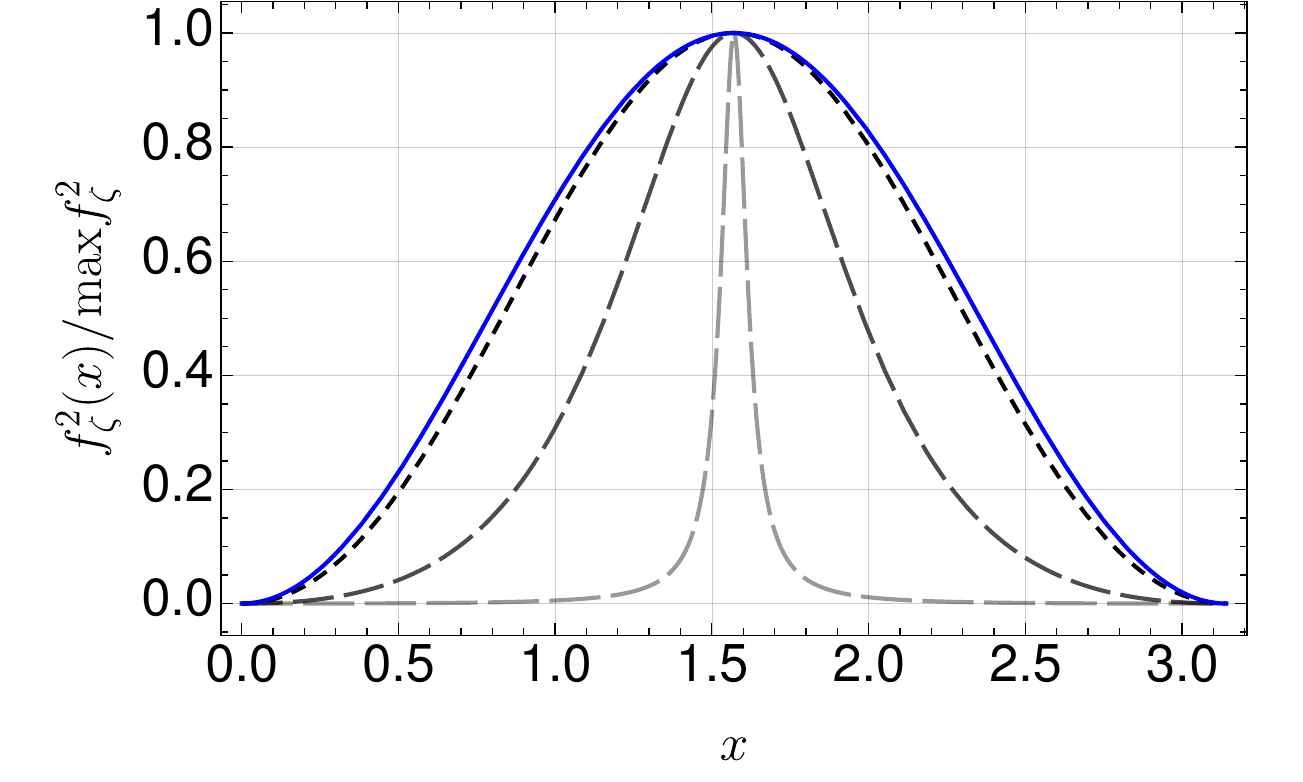}
\caption{Graphical representation of the $f_\zeta$ polymerization choice for various values of $\zeta$  ($\zeta=5$ in black with short dashes, $\zeta=1$ in gray with intermediate dashes and $\zeta=0.1$ in light gray with big dashes). The usual $\text{sin}^2(x)$ prescription is represented in solid blue line.}
\label{fig:fzeta}
\end{figure}

To specifically study the effects of the change of signature, we introduce a new function $f_{\zeta}(x)$ which has GR as a limit and allows to parametrically  control the $G^{(2)} \leq 0$ region. It reads:



\begin{equation}
    f_{\zeta}(x) = \sqrt{\frac{\zeta^2 + \pi^2}{\zeta^2 + 4\left( x - \pi/2\right)^2}} \sin(x),
    \label{eq:fzeta}
\end{equation}

where, $\zeta$ is the free parameter associated  with the signature change. Figure \ref{fig:fzeta} shows how the parameter changes the shape of the function.

\subsection{Tensor primordial power spectrum}

\begin{figure}[H] 
\includegraphics[width=\linewidth]{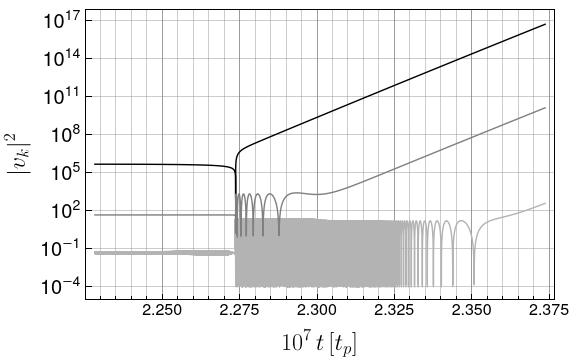}
\caption{Time dependence of the tensor mode amplitudes $|v_k|^2$, in the $f_\zeta$ polymerization framework, with $\zeta = 1$, for the comobile wave numbers $k=10^{-6}$, $k=10^{-2}$, and $k=10$ from top to bottom.}
\label{fig:temp}
\end{figure}

For illustrative purposes, the numerical computation of the time evolution of the tensor mode amplitude squared $|v_k|^2$ is displayed in Fig. \ref{fig:temp} for the $f_\zeta$ polymerization choice with $\zeta = 1$. As for other plots, Planck units are used. The contraction phase can easily be seen, together with the bounce, close to $t=2.275 \times 10^{7}~ t_{P}$, and the inflationary phase on the right side of the plot.

\begin{figure}[H] 
\includegraphics[width=\linewidth]{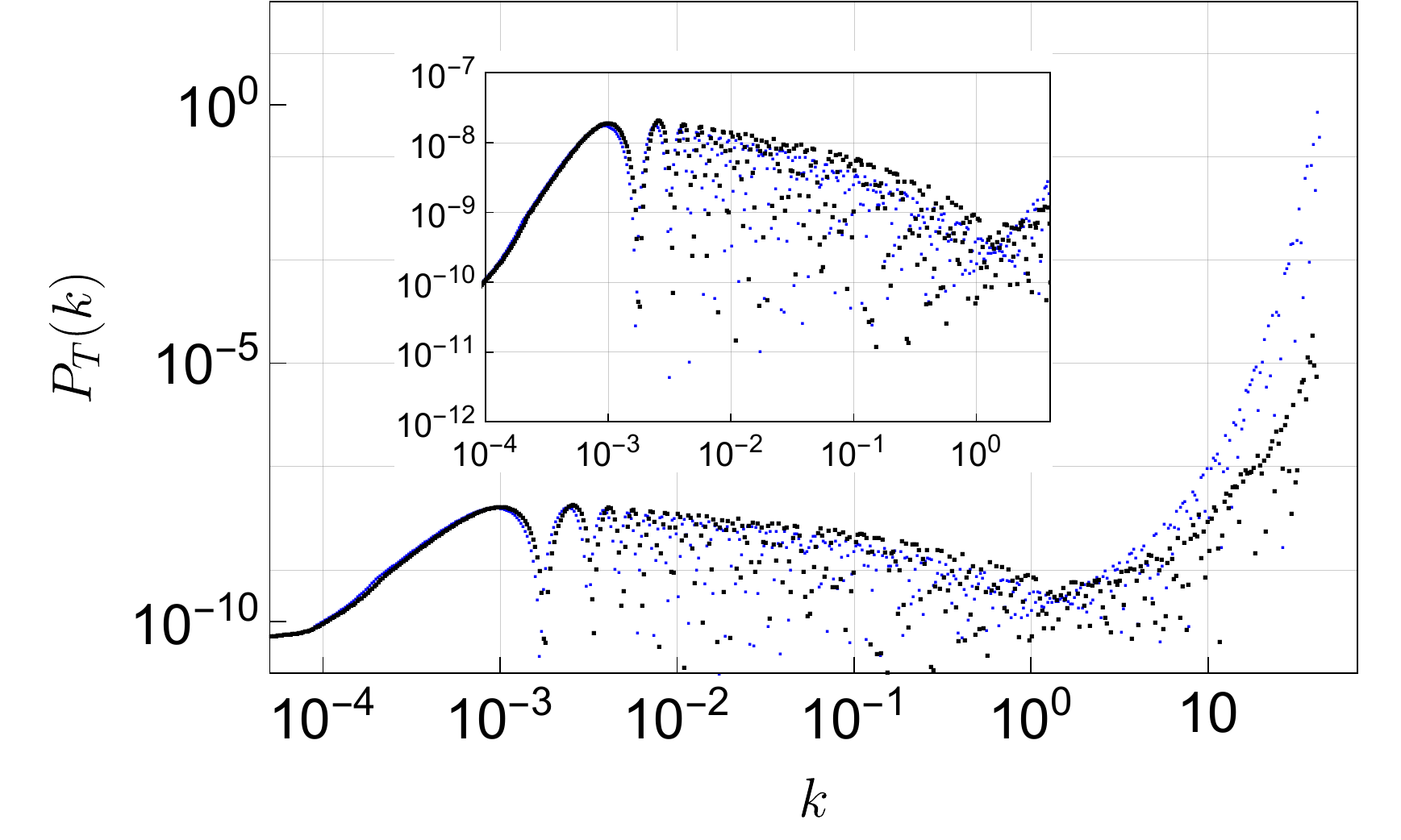}
\caption{Tensor primordial power spectrum for the $f_{\text{sqr}}$ polymerization (black dots) with $A_1=1$ and $n_i=1$ ($i\in\{1,2\}$), together with the usual $\text{sin}^2(x)$ prescription (smaller blue dots). A zoom on the oscillatory regime is also represented.}
\label{fig:ptfsq}
\end{figure}

\begin{figure}[H] 
\includegraphics[width=\linewidth]{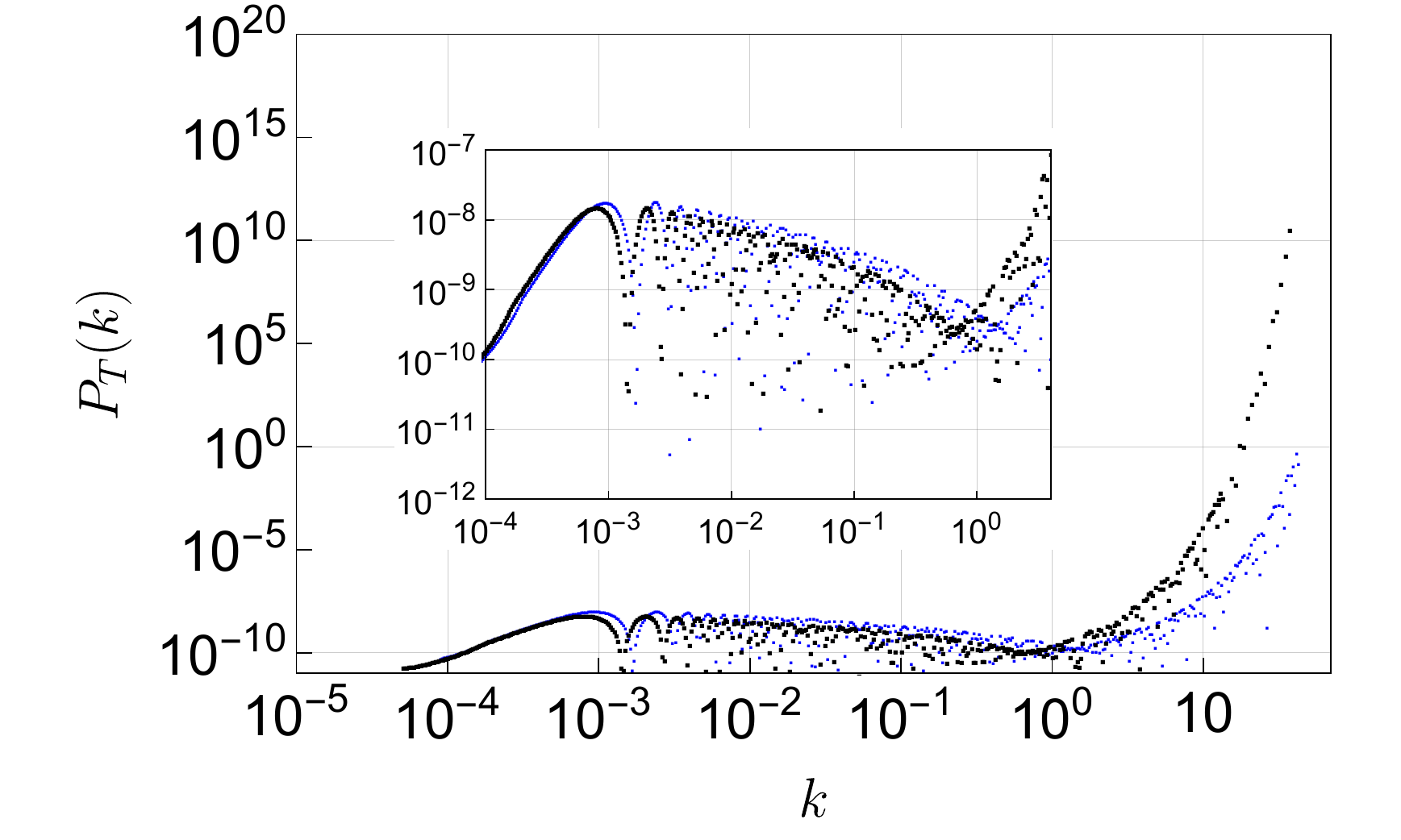}
\caption{Tensor primordial power spectrum for the $f_{\text{cos}}$ polymerization (black dots) with $C_1=2$, together with the usual $\text{sin}^2(x)$ prescription (smaller blue dots). A zoom on the oscillatory regime is also represented.}
\label{fig:ptfcos}
\end{figure}

\begin{figure}[H]
\includegraphics[width=\linewidth]{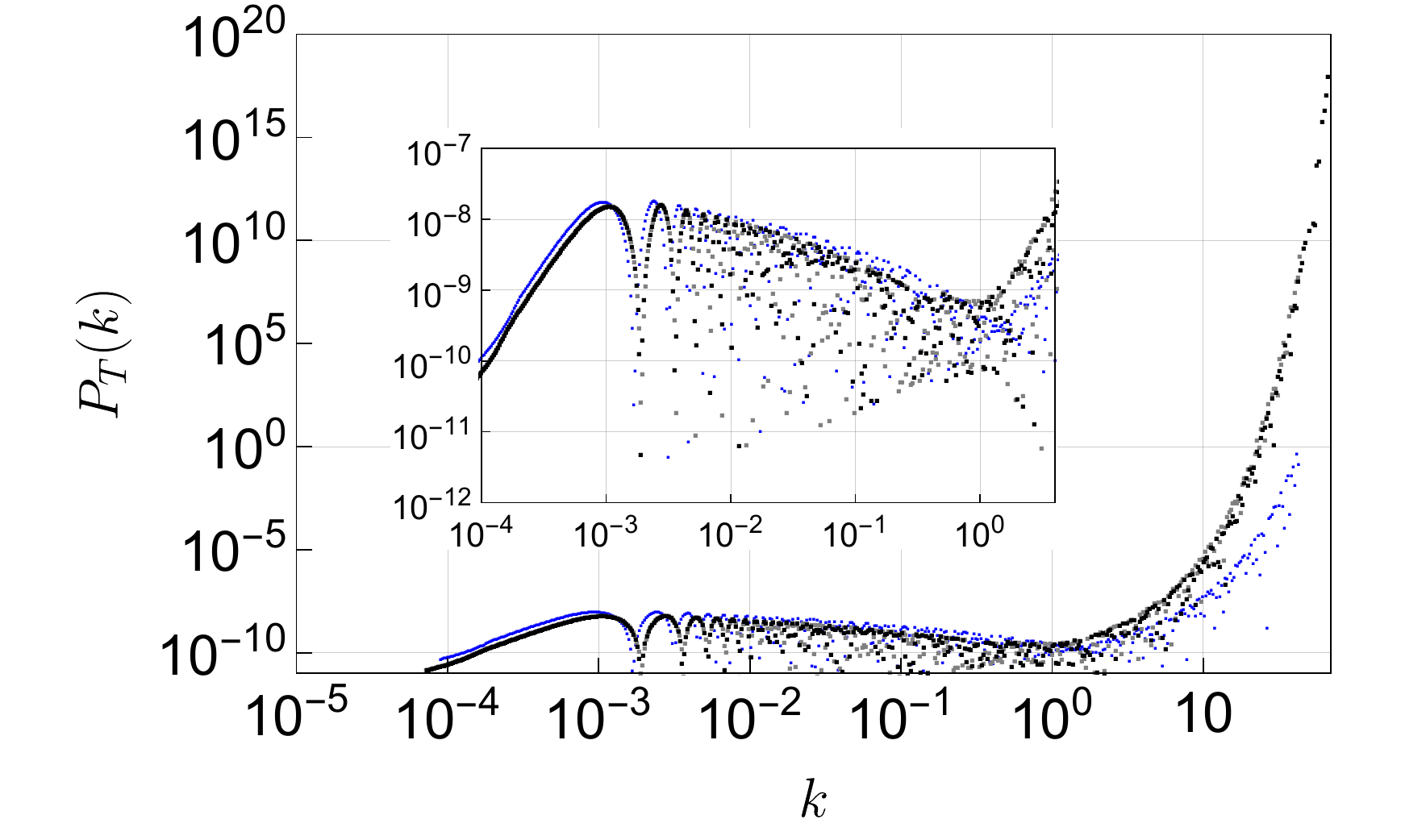}
\caption{Tensor primordial power spectrum for the $f_{\zeta}$ polymerization with $\zeta=1$ (black dots) and $\zeta=0.1$ (gray dots), together with the usual $\text{sin}^2(x)$ prescription (blue dots). A zoom on the oscillatory regime is also represented.}
\label{fig:ptfzeta}
\end{figure}

Tensor primordial power spectra for the polymerization choices defined in the previous section are presented in Figs. \ref{fig:ptfsq},\ref{fig:ptfcos}, and \ref{fig:ptfzeta}. Those spectra are evaluated at the end of the slow-roll inflationary phase, when slow-roll hypotheses break down. We ensure that all modes of interest, {\it i.e} those represented on the spectra, are  outside the horizon at that moment.
Whatever the polymerization choice, the spectra exhibit three regimes:

\begin{enumerate}
	\item A scale invariant behavior in the infrared limit $\left(k\in\left]-\infty; 10^{-4}\right]\right)$;
	\item An oscillatory behavior (associated with the bounce) for $\left(k\in\left[10^{-4};2\right]\right)$; and
	\item An exponential divergence in the ultraviolet limit $\left(k\in\left[2;+\infty\right[\right)$.
\end{enumerate}

The infrared (IR) regime corresponds to the largest wavelengths. Those modes do exit the horizon during the contracting phase (before slow-roll inflation takes places)  and remain frozen during the bounce and the subsequent phase of inflation. However, during contraction, the comoving Hubble radius (hence the tensor potential) behaves similarly as it does during the slow-roll phase $aH \sim 2 / \eta$ and $z_T''/z_T \sim 2 / \eta^2$. Moreover, the tensor potential converges towards 0 when going backward in time in the contracting branch and those modes can initially be normalized using the usual Bunch-Davies vacuum. This situation therefore corresponds to the classical solution of inflation and the associated tensor power spectrum is scale invariant. Those perturbations are not impacted by the presence of the bounce. As will be discussed further in this paper, the situation is more complicated for scalar perturbations as the scalar potential does not vanish when going backward in time in the contracting branch, leading to a scale-dependent behavior of the scalar spectra in the IR. This issue has been investigated in \cite{Barrau:2018gyz} but is however not of high importance as, for the vast majority of the parameter space, those modes cannot be observed in the CMB. A more exhaustive interpretation of those results has been widely studied in previous articles \cite{Linsefors:2012et,Bolliet:2015bka,Bolliet:2015raa,Barrau:2016sqp,Martineau:2017tdx}. It basically means that, depending on the number of inflationary e-folds, the model is either indiscernible from GR\footnote{We do not consider here the subtle normalization effects associated with the preceding deflation} (brief inflation), marginally compatible with GR, or fully different from GR (long inflation). We insist once more that the UV increase is not in itself inconsistent as the power spectrum does anyway {\it not} describe the real world in the $k \rightarrow \infty$ limit. Nonlinear local effects rule in this regime. In addition, it should be pointed out that modes are propagated in the Euclidean regime using their Fourier expansion which remains conceptually unclear.

\begin{figure}[H] 
\includegraphics[width=\linewidth]{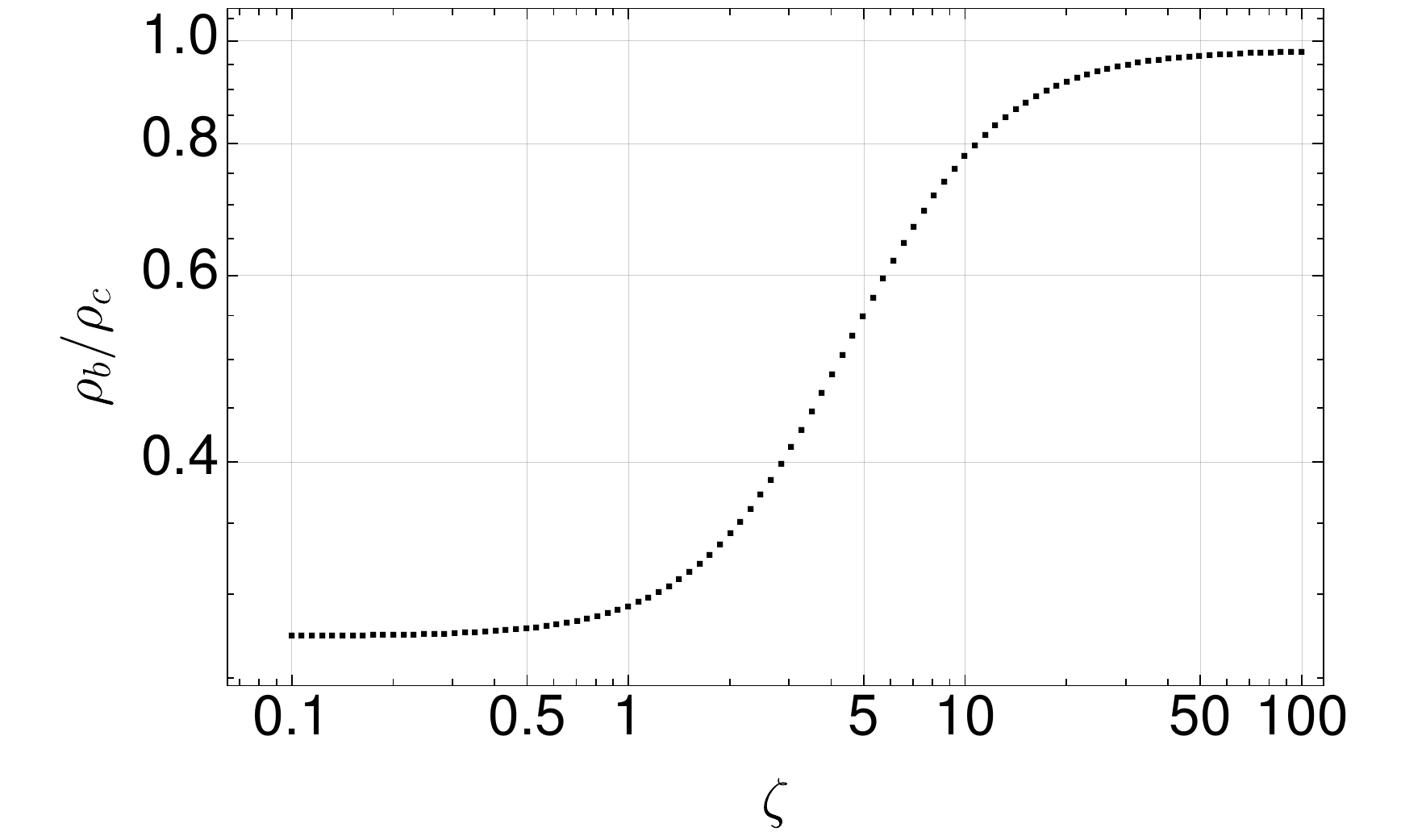}
\caption{$\zeta$ dependence of the bounce energy density for a background described by the $f_\zeta$ polymerization choice.}
\label{fig:rhob}
\end{figure}

The results displayed in the previously mentioned figures are not difficult to interpret (within the assumptions of the model). For tensor modes, the potential $z_T''/z_T$ depends only on the scale factor $a$ and its derivatives. In other words, the potential depends only upon background variables. Even with quite exotic generalized holonomy corrections, those variables are mostly equivalent to the usual loop quantum cosmology ones (see \cite{Renevey:2021tmh}). Nevertheless, some deviations from the standard behavior can be observed in the ultraviolet. This is due to the maximum value of $f$: if it differs from unity, the bounce energy density is not exactly the same than the one of the usual $\text{sin}^2(x)$ bounce. This can be explicitly seen in Fig. \ref{fig:rhob} for the $f_\zeta$ polymerization choice.

\subsection{Scalar primordial power spectrum}

\begin{figure}[H] 
\includegraphics[width=\linewidth]{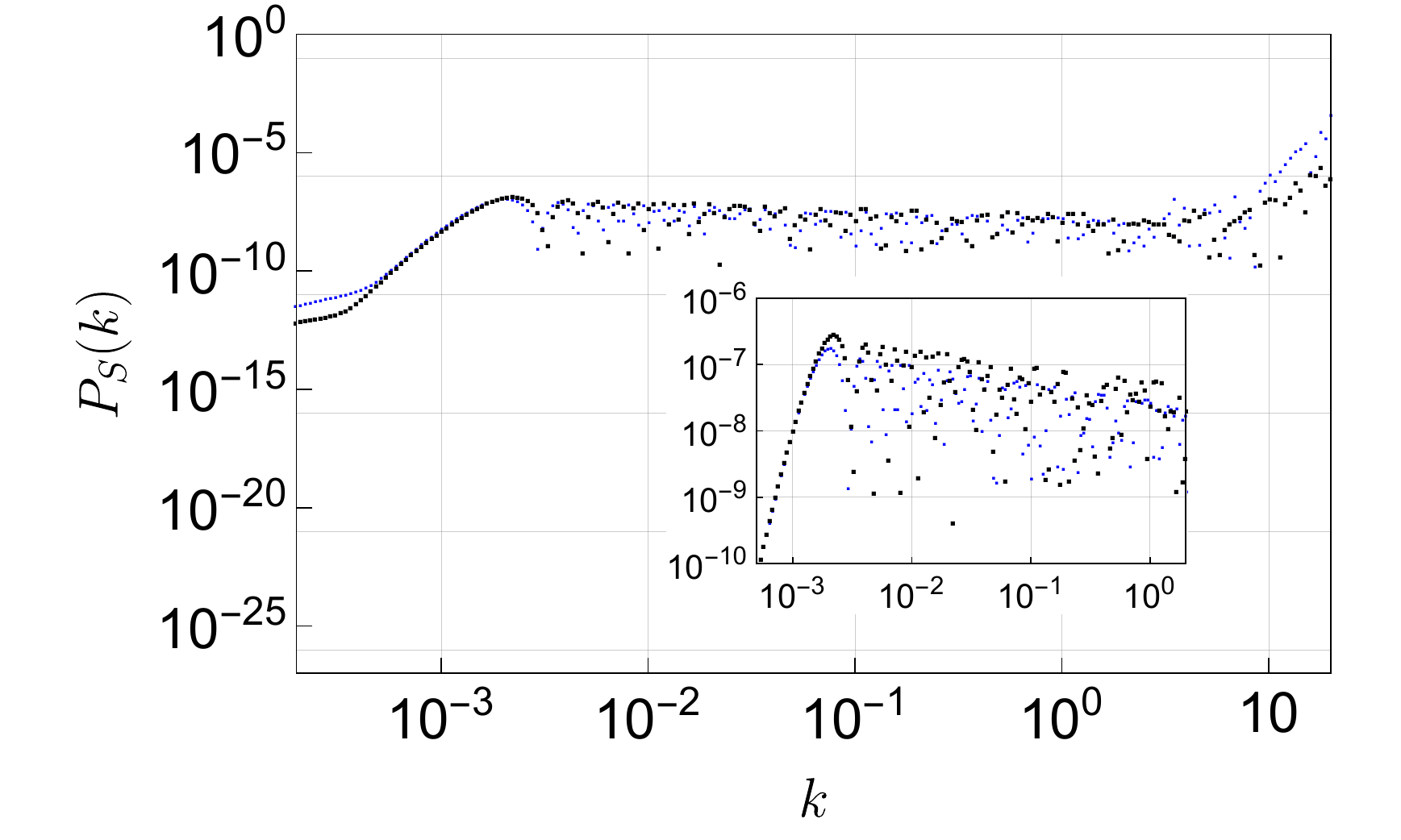}
\caption{Scalar primordial power spectrum for the $f_{\text{sqr}}$ polymerization (black dots) with $A_1=1$ and $n_i=1$ ($i\in\{1,2\}$), together with the usual $\text{sin}^2(x)$ prescription (blue dots). A zoom on the oscillatory regime is also represented.}
\label{fig:fsqrtnum}
\end{figure}

\begin{figure}[H] 
\includegraphics[width=\linewidth]{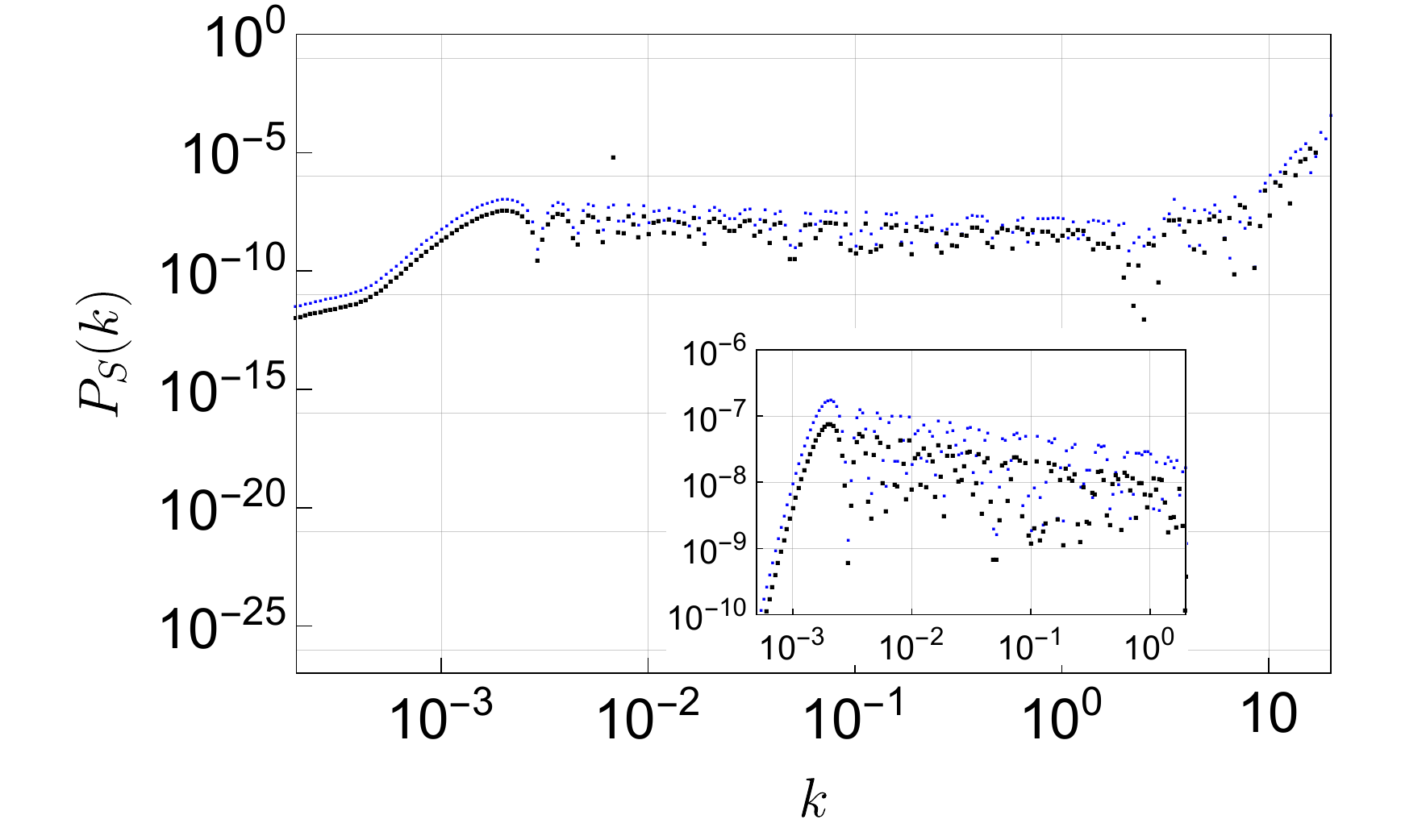}
\caption{Scalar primordial power spectrum for the $f_{\text{cos}}$ polymerization (black dots) with $C_1=2$, together with the usual $\text{sin}^2(x)$ prescription (smaller blue dots). A zoom on the oscillatory regime is also represented.}
\label{fig:fcosnum}
\end{figure}

\begin{figure}[H] 
\includegraphics[width=\linewidth]{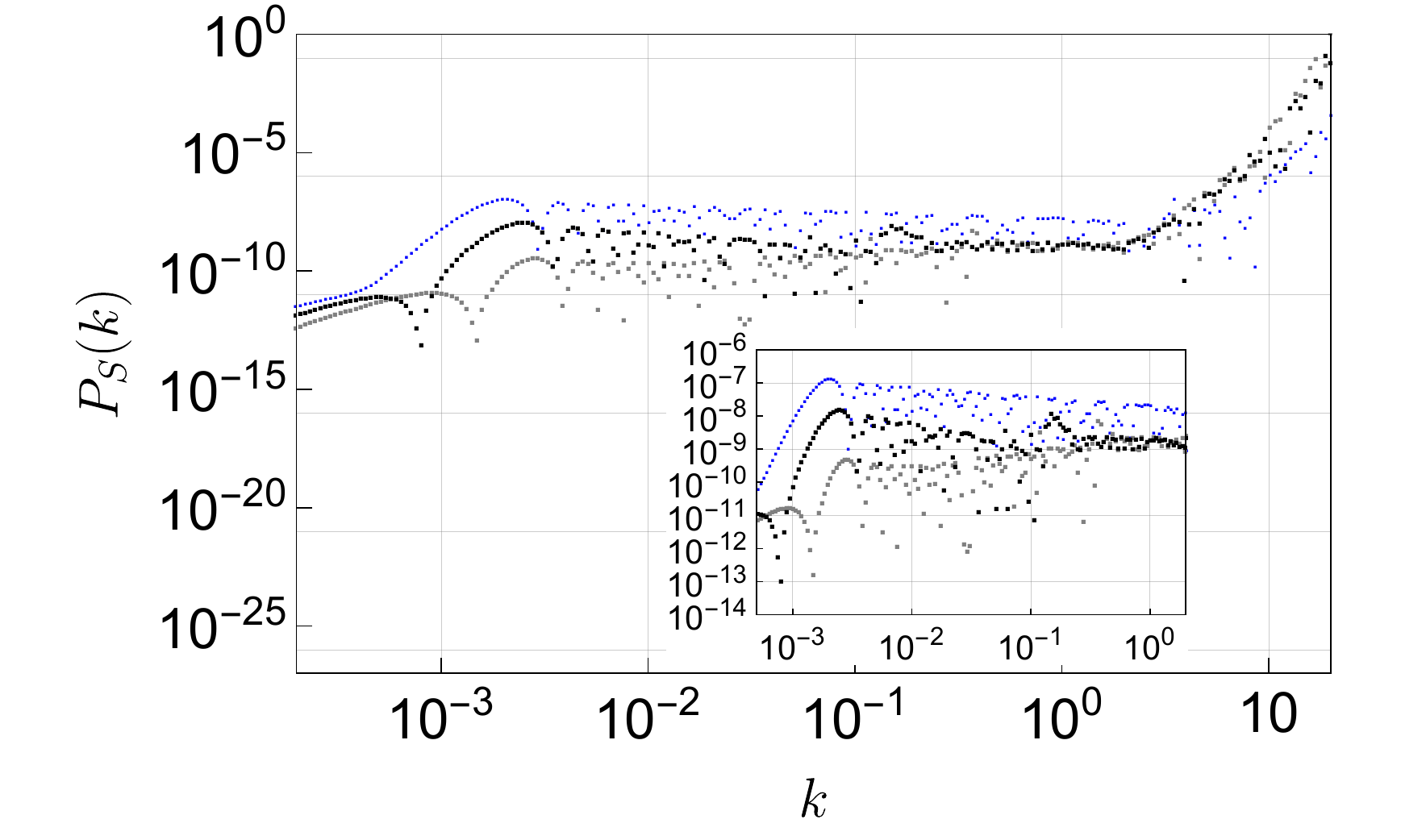}
\caption{Scalar primordial power spectrum for the $f_{\zeta}$ polymerization with $\zeta=1$ (black dots) and $\zeta=0.1$ (gray dots), together with the usual $\text{sin}^2(x)$ prescription (blue dots). A zoom on the oscillatory regime is also represented.}
\label{fig:fzetanum}
\end{figure}

The numerical results for the scalar primordial power spectra are given in Figs. \ref{fig:fsqrtnum},\ref{fig:fcosnum}, and \ref{fig:fzetanum} for the same polymerization choices. Three regimes can still be identified:

\begin{enumerate}
	\item A power law ($\propto k^3$) in the infrared  $\left( k\in\left]-\infty;10^{-3}\right]\right)$;
	\item Oscillations for $\left( k\in\left[10^{-3};2\right]\right)$;
	\item A divergence in the ultraviolet  $\left( k\in\left[2;+\infty\right[\right)$.
\end{enumerate}

Once again, the meanning of the main features have already been studied (see, in particular, \cite{Schander:2015eja}).

Starting from the definition of $z_S:=a^2 \dot{a}^{-1} \dot{\phi}$, one  obtains:

\begin{gather}
    \frac{z_S''}{z_S} = - a^2 \left( \partial^2_\phi V \right) + 2 \dot{a}^2 - 2 \kappa f'' \dot{\phi} \left( \partial_\phi V \right) a^4 \dot{a}^{-2} \\
    - \frac72 a^2 \kappa f'' \dot{\phi}^2 + 3 a^2 \kappa \dot{\phi}^4  + \frac12 a^4 \dot{a}^{-2} \kappa^2 f''^2 \dot{\phi}^4 \notag.
\end{gather}

 This complicated expression is what makes the case of scalar perturbations specific. In addition to the previously explained issues with initial conditions (that are not related with this specific model but inherent to all bouncing models), the fact that the polymerization choice also appears in the $ z_S''/z_S$ term of the propagation equation is what makes the study of scalar modes subtle. This is however the most interesting part of the game as they are directly related with CMB measurements.\\

To allow a direct comparison with data, one needs to convert comobile values into physical ones. In this work, as usual and useful when studying bouncing models, we normalized the scale factor to unity at the bounce time. The conversion therefore requires to know the number of e-folds between the bounce and the decoupling. In particular, it requires the knowledge of the number of inflationary e-folds $N_{inf}$, which cannot be fully fixed by the model but depends on contingent parameters (such as the phase of the scalar field during the contraction phase). Extensive discussions can be found in \cite{Bolliet:2015raa} and \cite{Barrau:2020nek,Renevey:2020zdj}. In practice, the physical wavenumber $k_{phys}$ is related to the comobile wavenumber $k$ used in the different plots of this article by
\begin{equation}
k_{phys}=k\left( e^{N_{inf}}\frac{T_{RH}}{T_{dec}}\right)^{-1},
\end{equation}
where $T_{RH}$ and $T_{dec}$ are, respectively, the reheating and decoupling temperatures.

The main conclusion that can be drawn from all the plots is that the spectra remain remarkably close one to the other, and similar to the ``standard" deformed algebra $\text{sin}^2(x)$ one. In the scalar case, this was not an {\it a priori} expected result. This shows that the precise shape of the holonomy correction has a very weak influence on the details of the observables, even if initial conditions are set in the contracting branch, and the perturbations propagated through the bounce and the Euclidean phase. This is an important point for the reliability of the model.

This work focuses on the {\it shape} of the power spectrum as this is precisely where effects of generalized holonomy corrections are expected to play a significant role. General considerations on the amplitude of the spectrum and on the scalar-tensor ratio in this framework can be found in \cite{Bolliet:2015raa,Barrau:2016sqp}.\\

\section{Falsifiability}

In principle, it could be that measurements allow to constrain the inflaton potential. In this case, the duration of inflation would somehow be predicted by the model (see, in \cite{Renevey:2021tmh}, the extension to generalized holonomy correction of the results from \cite{Linsefors:2014tna}). Should the latter be high enough so that the observational window falls in the UV part of the spectrum, the model would be discarded. What conclusions could then be drawn? Obviously, the situation is intricate as quite a few of explicit and implicit hypotheses are always at play in a cosmological scenario. At the heuristic level, this would discard a specific approach to generalizing gravity through a periodization of the Ashtekar connection (with the suitable low-curvature limit). More deeply, the delicate and important point would be to clarify the link between this specific approach and LQG in general.\\

On the one hand, it is true that the deformed algebra framework does not retain much of the complicated structure of loop quantum gravity. It particular, it is obviously ``less quantum" than the dressed metric approach.

On the other hand, it could be argued that it actually captures the core ingredients. Gauge fixing before quantization is often harmless. However, the constraints considered here are not of the kind of those encountered in Yang-Mills theories. When quantum corrected, the gauge transformation they generate are not of the usual form. Therefore, gauge fixing before quantization might lead to choose the gauge according to transformations that need to be modified; hence the inconsistency.

In addition, in the case of gravity, the dynamics is part of the gauge system \cite{Rovelli:2013fga,Rovelli:2020mpk}. Consistency therefore imposes to quantize gauge transformations and the dynamics simultaneously. It is not correct to fix the gauge in order to derive the dynamics. The deformed algebra approach solves both of those issues \cite{Barrau:2014maa} and should therefore be taken seriously.\\

The relation between this approach and the full theory is still unclear. If it was, in the future, shown to be reliably related to LQG, possible conflicts with data would rule out the main theory. On the other hand, if this framework was demonstrated to miss key features of the full theory, it would discard only this specific way to describe the cosmological dynamics.\\

It is fair to underline that this remains an open question at this stage. It should however be stressed that the unforeseen link between the deformed algebra approach and the disappearance of time, as predicted by the Hartle-Hawking proposal \cite{Hartle:1983ai}, is quite remarkable. Even more impressive is the way it might cure the weaknesses of the original proposal \cite{Bojowald:2018gdt,Bojowald:2020kob}.\\

Finally, as previously reminded, it could also be that modes are not correctly propagated in the Euclidean phase. In this work, we make minimalist assumptions and work in Fourier space to avoid obvious problems with the definition of a wave in a timeless space. Another interesting view was suggested in \cite{Bojowald:2015gra}.

\section{Conclusion}

In this work we have considered generalized holonomy corrections, as the usual harmonic choice made in loop quantum cosmology is far from being the only possible one. It has even been recently argued that there is no fundamental reason for focusing on this specific shape \citep{Amadei:2022zwp}. 

We have studied three different generic functions having general relativity as their low-energy limit and satisfying the basic loop gravity requirements. One of them is specifically parametrized so that the position of the Euclidean region, corresponding to a change of concavity, can be easily varied and probed.\\

The generalized holonomy correction appears both at the background level and in the propagation equation for perturbations. In addition, for scalar modes, it also enters the $z''_S/z_S$ term. This leads to an intricate situation which cannot be fully understood intuitively.

To clarify the situation, we have numerically calculated the primordial power spectra in all cases, setting initial conditions in the prebounce contracting branch. Since in this setting (motivated by general arguments), cosmological perturbations are propagated through the bounce and the Euclidean phase, a bigger sensitivity of the spectra to the shape of the holonomy correction than the one established in \citep{Han:2017wmt} could have been expected. However, we have shown that, whatever the (reasonable) form of the function and values of the parameters, the overall shape of the spectra remains unchanged with respect to the usual deformed algebra LQC results. This shows that the known conclusions are robust.\\

Obviously, the actual content of the Universe in the contraction phase is not known and this constitutes a weakness for all bouncing models. As pointed out in \cite{Barrau:2014kza,Barrau:2017ukm} this might raise some interesting paradoxes. In this work, the only assumption required is that a scalar field dominates over all the other possible contents at high-energy before the bounce. Although speculative, this assumption makes sense as it both leads to the desired phase on inflation and seems favored by grand unified models of particle physics \cite{Mohanty:2008ab}. Obviously, a detailed description on an ``inverse-reheating" process is still missing. More important than the actual content is the question of anisotropies, extensively discussed {\it e.g.} in \cite{Martineau:2017sti}.\\

In the future, it would be interesting to generalize this investigation to the dressed metric approach. In this case, the way the new holonomy correction might alter the propagation equation is, however, less clear and requires further investigations.

\bibliography{refs.bib}

 \end{document}